\documentclass[onecolumn,preprintnumbers,amsmath,amssymb]{revtex4}
\usepackage{graphicx}
\usepackage{dcolumn}
\usepackage{bm}
\def\be{\begin{equation}}
\def\ee{\end{equation}}
\def\bea{\begin{eqnarray}}
\def\eea{\end{eqnarray}}

\begin{document}

\title{On the measurement of the muon anomalous magnetic moment}

\author{Ara Ioannisian}

\affiliation{ 
Yerevan Physics Institute, Alikhanian Brothers 2, Yerevan-36, Armenia\\
 Institute for Theoretical Physics and Modeling, Yerevan-36, Armenia }

\begin{abstract}
The ideas and formulas presented in the article will help to bring together the theoretical predictions 
for the anomalous magnetic moment of muon and the results of the "Muon g-2" experiment. In doing so, 
we are discussing the new effect exclusively within the Standard Model.
In quantum physics a state with spin perpendicular to a magnetic
field can be expressed as a superposition of energy eigenstates with
spins parallel and  antiparallel to the field: the resultant spin
precession is due to the energy difference between the two eigenstates.
If the state, like the muon, is unstable and can decay, it will
have a natural energy spread. As a result the frequency of the spin
precession can vary.  For a constant magnetic field the measured spin
precession velocity will be spread according to the Lorentzian
distribution with width $\left(\gamma\tau\right)^{-1}$, for Lorentz
gamma factor $\gamma=E/ m$, and particle lifetime $\tau$. Although the
true mean and variance of a Lorentzian distribution is undefined, the
latter can be estimated by the maximum likelihood method to be ${2
  \over N (\gamma \tau)^2}$, twice that of a normal distribution. 
 Thus, the statistical error on the anomalous magnetic moment in reality
 should turn out to be wider than with $\chi^2$ analysis of the experiment.
\end{abstract}
\maketitle

Recently Fermilab's ``Muon g-2'' collaboration announced new results on
the measurement of the muon anomalous magnetic moment
\cite{Muong-2:2021ojo}.  Their measurement, which supports the
previous result by the E821 experiment at Brookhaven National
Laboratory \cite{Muong-2:2006rrc},  differs significantly from the
theoretical prediction of the muon anomalous magnetic moment.

The physics of spin precession in a magnetic field is well understood
within classical relativistic physics, see e.g. \cite{UFN-Malykin,
  Bargmann:1959gz}.  However the muon in an unstable particle and as
such it has a natural width, which can only be treated correctly within quantum physics.  

When in the magnetic field a charged fermion has two independent
energy eigenstates with spin parallel and anti parallel to the
field. Those two states gain energy with opposite signs. Due to that
energy difference ($\Delta E$) the direction of the spin,
perpendicular to the magnetic field, will begin to precess.   

\bea
\psi^\rightarrow ={1 \over \sqrt{2}} (\psi^\uparrow+ \psi^\downarrow)\ , \ \ \ 
\psi^\leftarrow ={1 \over \sqrt{2}} (\psi^\uparrow- \psi^\downarrow)
\\
\psi(t) ={1 \over \sqrt{2}} e^{-iE^\uparrow t}(\psi^\uparrow+e^{i(E^\uparrow-E^\downarrow)t} \psi^\downarrow)={1 \over \sqrt{2}} e^{-iE^\uparrow t}(\psi^\uparrow+e^{i \Delta E t} \psi^\downarrow)
\eea
\bea
P_{\psi_\rightarrow \Rightarrow \psi_\rightarrow}= \cos^2 {\Delta E \ t \over 2 }
\\
P_{\psi_\rightarrow \Rightarrow \psi_\leftarrow}= \sin^2 {\Delta E \ t \over 2 }
\eea
The experiment (FNAL/BNL) is measuring spin precession relative to the muon velocity direction. In that frame
\be
\Delta E_0 ={g-2 \over 2 }{e B \over  m} =a_\mu {e B \over m} \ \ \ (a_\mu={\alpha \over 2 \pi}+ \dots)
\ee

In case the fermion is an unstable particle  the energy of each of the
eigenstates (with spins parallel and antiparallel to the magnetic field) will have its own energy spread according to the Lorentz distribution
\be
{1 \over \pi \ 2 \gamma \tau} {1\over (E^i-E^i_0)^2+{1\over 4 (\gamma \tau)^2}}  \ , \ \ \ i=\{ {\rm spin} \uparrow, \  {\rm spin} \downarrow \}
\ee

The energy difference $\Delta E = E^\uparrow - E^\downarrow$ will also
have a Lorentz distribution with double the width 
 \be
{1 \over \pi \ \gamma \tau} {1\over (\Delta E-\Delta E_0)^2+{1\over (\gamma \tau)^2}}  \ , \ \ 
\ee

This additional energy difference arises during entrance of the unstable fermion into the magnetic field.

There is no expectation and  no  standard deviation for  the Lorentzian distribution; the integrals diverge \cite{VanderWaerden}. 
The $\chi^2$ analyses of the data is not defined.  For large $N$ ($N$ is the number of measurements) asymptotically the maximum
 likelihood method predicts the location of the maximum
($\Delta E_0$) and its variance $\sigma^2_{\Delta E_0}= {2 \over N
  (\gamma \tau)^2}$ \cite{VanderWaerden}.

This additional factor $2$, which appears due to the Lorentz distribution, is absent in case of normal distribution. This factor will increase 
the statistical error of the experiment and accordingly expand the possibilities of convergence between theory and experiment. 

We are grateful to the staff of the Institute of Physics of Zurich University for the warm hospitality  and 
acknowledge the contribution of the COST Action CA18108 “Quantum gravity phenomenology in the multi-messenger approach”. 
We would like to thank G. Giudice, Rakhi Mahbubani, G.Isidori, V.Mekhitaryan and J.Kopp for discussions.

\end{document}